\documentclass{vgtc}                          

\ifpdf
  \pdfoutput=1\relax                   
  \usepackage{graphicx}                
  \DeclareGraphicsExtensions{.pdf,.png,.jpg,.jpeg} 
\else
  \usepackage{graphicx}                
  \DeclareGraphicsExtensions{.eps}     
\fi%

\graphicspath{{figures/}{pictures/}{images/}{./}} 

\usepackage{microtype}                 
\PassOptionsToPackage{warn}{textcomp}  
\usepackage{textcomp}                  
\usepackage{mathptmx}                  
\usepackage{times}                     
\usepackage{cite}                      
\usepackage{tabu}                      
\usepackage{booktabs}                  
\usepackage{color}
\usepackage{multirow}
\usepackage{xspace}
\usepackage[normalem]{ulem}
\usepackage{setspace}
\usepackage{color,soul}
\definecolor{lightgray}{gray}{0.75}
\sethlcolor{lightgray}

\onlineid{0}

\vgtccategory{Research}

\vgtcinsertpkg

\title {You Can't Publish Replication Studies (and How to Anyways)\\ \vspace{1ex} \large Position Paper}

\author{Ghulam Jilani Quadri\thanks{e-mail: ghulamjilani@mail.usf.edu}\\ %
        \scriptsize University of South Florida %
\and Paul Rosen\thanks{e-mail: prosen@usf.edu}\\ %
     \scriptsize University of South Florida %
}

\abstract{
Reproducibility has been increasingly encouraged by communities of science in order to validate experimental conclusions, and replication studies represent a significant opportunity to vision scientists wishing contribute new perceptual models, methods, or insights to the visualization community.  Unfortunately, the notion of replication of previous studies does not lend itself to how we communicate research findings. Simple put, studies that re-conduct and confirm earlier results do not hold any novelty, a key element to the modern research publication system. Nevertheless, savvy researchers have discovered ways to produce replication studies by embedding them into other sufficiently novel studies. In this position paper, we define three methods---re-evaluation, expansion, and specialization---for embedding a replication study into a novel published work. Within this context, we provide a non-exhaustive case study on replications of Cleveland and McGill's seminal work on graphical perception. As it turns out, numerous replication studies have been carried out based on that work, which have both confirmed prior findings and shined new light on our understanding of human perception. Finally, we discuss how publishing a true replication study should be avoided, while providing suggestions for how vision scientists and others can still use replication studies as a vehicle to producing visualization research publications.
} 

\begin{document}

\firstsection{Introduction}
\maketitle

A replication study is a re-evaluation, re-confirmation, or extension of an original study. These studies can be performed under similar experimental conditions to the original studies to validate conclusions or under varying experimental conditions to gain more knowledge~\cite{peng2009reproducible, hornbaek2014once}, and they provide an important concrete baseline, which is useful to improving cross-study validity~\cite{lallemand2013replicating}. The journal \textit{Science} labeled replication as the ``scientific gold standard''~\cite{jasny2011again} and in others' words ``in replication, the private chimera becomes the communal fact''~\cite{BAZERMAN}.

Unfortunately, the positivist notion of reproducibility does not offer the novelty of qualitative research itself~\cite{sukumarreplication}. In a literature review of 891 papers by Hornbaek et al., they found only 3\% of papers in Human-Computer Interaction (HCI) attempted replication~\cite{hornbaek2014once}, and at a recent BELIV workshop multiple researchers pointed out that it is rare to find replication work in information visualization research~\cite{kosara2018skipping,sukumar2018towards}.

Efforts have been made to encourage researchers to improve reproducibility and conduct more replication studies. 
For example, it is increasingly common for authors to voluntarily (and sometimes mandatorily) provide their experimental data and other supplemental materials to accommodate future replication. Since 2012, the EuroRVVV\footnote{\url{https://eagereyes.org/blog/2018/eurorvvv-call-for-papers}} Workshop on Reproducibility, Verification, and Validation in Visualization has been promoting and supporting replication in visualization. The RepliCHI\footnote{\url{https://replichi.org}} workshop, organized as a part of the ACM CHI conference, promotes a shift towards favoring replication in the HCI research community~\cite{wilson2014replichi}. Finally, the BELIV\footnote{\url{https://beliv-workshop.github.io}} workshop at IEEE VIS conference is also focused on {replication} and {reproducibility}~\cite{valdez2018requirements,sukumar2018towards,kosara2018skipping}.

A highly debated usability evaluation paper in ACM CHI highlighted the lack of replication studies in HCI~\cite{greenberg2008usability}. It argued that \textit{reviewers do not value replication work}, and therefore papers are not published, despite the intrinsic value in their work. The reluctance from the reviewers and publication venues to consider replication studies as a valuable contribution has lead to few replication studies being published~\cite{hornbaek2014once, wilson2014replichi, greenberg2008usability}. Most of the replications studies successfully published are of controversial findings or of highly-cited works~\cite{bronstein1990publication, greenberg2008usability, cleveland1984graphical, hornbaek2014once, jones2010investigation, rosenthal1990replication, dragicevic2017blinded}. Instead of attempting replication studies, HCI and information visualization researchers have been pushed towards novel and organic findings~\cite{newman1994preliminary, hornbaek2014once, wilson2014replichi}. Nevertheless, savvy researchers have been successful at publishing replication studies in less conventional ways, in particular, by embedding them into other sufficiently novel studies. Many of these studies have been at the intersection of vision science and visualization.

Recent workshops and conferences on vision science and visualization have established the potential for methodologies, experimental techniques, and user studies on perceptual judgements from vision science to benefit visualization. However, cross-field contribution is challenging. For vision scientists wanting to interact with the visualization community, \textit{replication} can be a viable and relatively low-risk area where they can contribute.

In this position paper, we highlight how researchers have integrated replications of prior studies into new studies by including them with cutting-edge contributions. In this way, these new works confirm the prior studies and introduce a novel idea or methodology. In particular, we found three common methods of integration: 
(1)~\textit{re-evaluating} under different demographics and/or participant environments;
(2)~\textit{expanding} upon a study's conclusions by new experiments that elucidate additional information or deepen understanding; or 
(3)~\textit{specializing} the knowledge to a specific domain by elaborating on experimental conclusions. 
To demonstrate these three methods at the intersection of vision science and visualization, we highlight a number of replicated works of Cleveland and McGill's graphical perception paper~\cite{cleveland1984graphical}, whose objectives are similar to the original ranking of quantitative judgment effectiveness of graphical encodings. We review their innovative contributions and contribution to replication-based validation. Finally, we discuss our perspective on how vision scientists can use this information for producing replication works in the current publication environment.

\section{Background}

The visualization field is ``an empire built on sand'' with a weak foundation and in need of replication to strengthen many of the assumptions used by the field~\cite{kosara2016empire}. In a recent study, Kosara and Haroz pointed out the \textit{replication crisis}---very few replication studies are attempted in visualization, let alone published~\cite{kosara2018skipping}. They examined six threats to the validity of studies in visualization and provided suggestions for replications, like outlining study design flaws and understanding or re-running misinterpreted results. Their suggestions help to minimize the threats to producing scientifically sound work.

Sukumar et al.~provided guidelines for experimental design to encourage researchers to conduct more replication studies in information visualization~\cite{sukumar2018towards}. They provided a list of possible experimenter biases that can occur related to devising hypotheses, independent and dependent variables, tasks, experimental procedures, sampling, experimenter behavior, and experimental setting, and they focused their discussion on designing and running sound experiments.

Hornbaek et al.\ developed a prescriptive definition for replication studies~\cite{hornbaek2014once}. They stated that a replication must name and reference the original work, and they must state how their work confirms or extends the prior study. They further distinguished the replicated work with three categories---\textit{strict}, \textit{partial}, and \textit{conceptual}---based on their literature review. A similar type of distinguished category can be found in Kosara and Haroz work---\textit{reanalysis}, \textit{direct replication}, \textit{conceptual replication}~\cite{kosara2018skipping}. These characterizations vary by the amount of originality from previous work that has been kept intact. \textit{Strict replication} uses the same variables with the intent to reproduce the exact same study. Strict replications usually only replicate and confirm the original findings, which is what draws reviewers reluctance. \textit{Partial replication} modifies the original study for testing within a different environment or with different participants. \textit{Conceptual replication} studies investigate the same study but with different metrics, settings, or judgment criteria.

We distinguish our contribution from previous studies~\cite{kosara2018skipping, hornbaek2014once, sukumar2018towards} by focusing on a taxonomy that does not focus on the quantity of overlap or similarity of study design in replication studies, but instead it focuses on the \textit{types of novel contributions} that are associated with the replication studies.

\section{Novelty and Replication}
 
Hornbaek et al.'s work~\cite{hornbaek2014once}, along with the Kosara and Haroz work~\cite{kosara2018skipping}, essentially considered the level of similarity to the original study when classifying replication. As a complementary measure to that, we consider classifying replication studies by the \textit{objective} of the novel contributions of the study. By understanding the types of novel contributions that blend with replication studies, other researchers, like those from vision science, can mimic the contribution styles to produce their own replication studies.

\subsection{Taxonomy of Replication}

    Our evaluation of prior work shows that the vast majority of replicated work in information visualization falls within one of the three following categories.

\subsubsection{Re-evaluate an Experiment's Objective}

As practices change, software and hardware advance, and new techniques or information become available, some research studies whose conclusions were once considered solid require re-evaluation using these new contexts. Re-evaluation studies confirm the findings of an original study with different environment setting, while attempting to reproduce the objectives as closely as possible~\cite{hornbaek2014once}. These replications help to establish if results from the prior study can be repeated to increase confidence in its validity.  For example, Kosara and Ziemkiewicz replicated~\cite{kosara2010mechanical} their own earlier studies~\cite{skau2016arcs,kosara2016judgment} on pie charts using Amazon Mechanical Turk in order to re-evaluate whether an online environment produced the same results. Additional examples of this type of replication study are~\cite{kay2015beyond,heer2010crowdsourcing}.

\subsubsection{Expand an Experiment's Objective}

Due to the efforts required for performing human studies, the scope of studies is often kept small, leaving the need to expand conclusions with followup studies. Replication studies can be expanded beyond the objectives and conclusions of the prior studies by conducting themselves under different experimental conditions to make a novel contribution. By experimenting under different conditions, these replication studies serve as alternative means to validate prior results or as a means to generalize results~\cite{jakobsen2013interactive, harrison2014ranking, kong2010perceptual}. For example, Rensink and Baldridge investigated the perception of correlation in scatterplots and suggested that Weber's law was useful for modeling it~\cite{rensink2010perception}. A replication study by Harrison et al.\ supplemented new conditions, in order to broaden the scope into a number of additional visualization types~\cite{harrison2014ranking} (which was itself extended further through re-evaluation type paper~\cite{kay2015beyond}). Some additional examples of these types of replication studies are~\cite{ottley2015improving, kim2012does,reda2018graphical}.

\subsubsection{Specialize an Experiment's Objective}

Studies often result in generalizable conclusions that need to be studied in new or more specific contexts than those of the original study. By taking the original study as the fundamental base, these works consider if and how much of the knowledge acquired in the original study is transferable to a new, different, or more specific domain. Performing these replication studies of different contexts under similar conditions with respect to setting, experiment environments, or metrics, aims to establish the validity within that specialized domain. Rensink and Baldridge's~\cite{rensink2010perception} study on modelling correlation perception was also replicated by Yang et al.~\cite{yang2018correlation} in order to further investigate visual features in modelling perceptual processes. The objective of finding perceived correlation in a scatterplot is synonymous with perceiving its visual features and quite unrelated to one's statistical training. The results of this replication study showed how visual features provide a baseline for model-approaches in visualization evaluation and design. Additional examples of this type of replication study can be found in~\cite{szafir2018modeling, heer2009sizing}.

\begin{table*}[!t]
    \centering
    \caption{Selected publications that are representative replication studies of Cleveland and McGill~\cite{cleveland1984graphical}. }
    \label{tbl:List_papers}
    \resizebox{1.0\linewidth}{!}{%
     \setlength{\tabcolsep}{10pt} 
     \def\arraystretch{1.35}
    \begin{tabular}{p{7.65cm}||cc|cc|c}
         \multirow{2}{*}{Publication} & 
          \multirow{2}{*}{Venue} &
         \multirow{2}{*}{Year} &
        Hornbaek et al.'s& 
        Replication & 
        Types of Graphical \\
        & & & Category~\cite{hornbaek2014once} & Type & Perception (from \autoref{fig:cm_encoding}) 
        \\
         
        \hline
        \hline
            Crowdsourcing Graphical Perception: Using Mechanical  & 
            \multirow{2}{*}{ACM CHI} & 
            \multirow{2}{*}{2010} & 
            \multirow{2}{*}{Conceptual} & 
            \multirow{2}{*}{\textbf{Re-evaluate}} & 
            \multirow{2}{*}{Length, Position \& Angle} \\ 
            Turk to Assess Visualization Design \cite{heer2010crowdsourcing} & & & & \\
         Evaluating Interactive Graphical Encodings for Data Visual- & 
            \multirow{2}{*}{IEEE TVCG} & 
            \multirow{2}{*}{2018} & 
            \multirow{2}{*}{Conceptual} & 
            \multirow{2}{*}{\textbf{Re-evaluate}} & 
            {12 encodings} \\
            ization \cite{saket2018evaluating} & & & & & (interactive encodings)  \\
          Graphical Perception of Continuous Quantitative Maps:  & 
            \multirow{2}{*}{ACM CHI} & 
            \multirow{2}{*}{2018} & 
            \multirow{2}{*}{Conceptual} & 
            \multirow{2}{*}{\textbf{Re-evaluate}} & 
            \multirow{2}{*}{Color} \\
            The Effect of Spatial Frequency and Color Map Design \cite{reda2018graphical} & & & & \\
         \hline
            The Impact of Social Information on Visual Judgments \cite{hullman2011impact} & 
            \multirow{1}{*}{ACM CHI} & 
            \multirow{1}{*}{2011} &
            \multirow{1}{*}{Partial} & 
            \multirow{1}{*}{\textbf{Expand}} &
            \multirow{1}{*}{Length, Position \& Area} \\ 
            Influencing Visual Judgment through Affective Priming~\cite{harrison2013influencing}  & 
            \multirow{1}{*}{ACM CHI} & 
            \multirow{1}{*}{2013} & 
            \multirow{1}{*}{Conceptual} & 
            \multirow{1}{*}{\textbf{Expand}} & 
            \multirow{1}{*}{Length, Position \& Angle} \\
          Learning Perceptual Kernels for Visualization Design \cite{demiralp2014learning}   & 
            \multirow{1}{*}{IEEE TVCG} & 
            \multirow{1}{*}{2014} & 
            \multirow{1}{*}{Partial} & 
            \multirow{1}{*}{\textbf{Expand}} & 
            \multirow{1}{*}{Shape, Color \& Size} \\
         \hline
          Sizing the Horizon: the Effects of Chart Size and Layering  & 
            \multirow{2}{*}{ACM CHI} & 
            \multirow{2}{*}{2009} & 
            \multirow{2}{*}{Partial} & 
            \multirow{2}{*}{\textbf{Specialize}} & 
            \multirow{2}{*}{Chart Height \& Layering} \\
             on the Graphical Perception of Time Series Visualizations~\cite{heer2009sizing} & & & &   \\
        
            \multirow{2}{*}{Four Experiments on the Perception of Bar Charts \cite{talbot2014four}}  & 
            \multirow{2}{*}{IEEE TVCG} & 
            \multirow{2}{*}{2014} & 
            \multirow{2}{*}{Partial} & 
            \multirow{2}{*}{\textbf{Specialize}} & 
            Length \& Position \\
            & & & & & (for bar charts) \\
           
    \end{tabular}} 
\end{table*}

\section{Case Study}

With a focus on perceptual work in information visualization, we selected the seminal and highly cited work by Cleveland and McGill on graphical perception~\cite{cleveland1984graphical} to highlight examples of replication studies. Graphical perception is an anchor of visualization research~\cite{ware2012information} with the potential to improve the efficiency and effectiveness of the automatic representation of data~\cite{mackinlay1986automating}. We have observed the vast influence of this paper in last two decades of information visualization research. According to Google Scholar, this work has been cite more than 1585 times as of July 1, 2019, and aspects of their study have been replicated many times.

\begin{figure}[!b]
    \centering
    \includegraphics[width=0.975\linewidth]{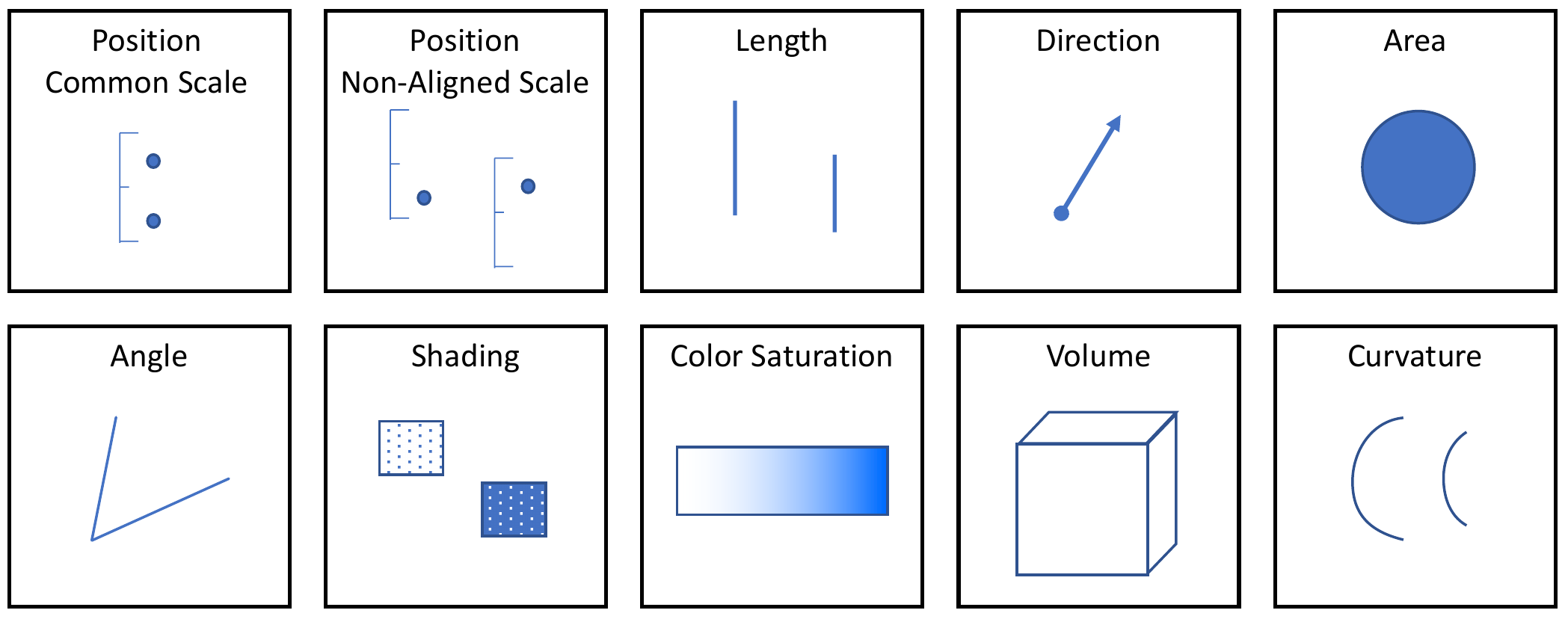}
    \caption{Reproduction of the Cleveland and McGill's graphical encoding channels~\cite{cleveland1984graphical}.}
    \label{fig:cm_encoding}
\end{figure}

\subsection{Graphical Perception}

Research has demonstrated viewer's perceptual judgment significantly influences effective visualization design. The representation of data in a visualization is encoded with specific elements on the display, also known as the graphical encodings, include \textit{position}, \textit{length}, \textit{angle}, \textit{area}, \textit{volume}, \textit{shading}, \textit{direction}, \textit{curvature}, and \textit{color}~\cite{pinker1990theory}. \autoref{fig:cm_encoding} represents these 10 elementary \textit{perceptual tasks}\footnote{The term \textit{perceptual task} originates from the concept that  viewer performs a mental task to extract quantitative values represented on graphs~\cite{cleveland1984graphical}.} from Cleveland and McGill's work that people use to extract quantitative information from graphs.

\begin{figure}[!b]
    \centering
    \includegraphics[trim=0 5pt 40pt 15pt, clip, width=0.975\linewidth]{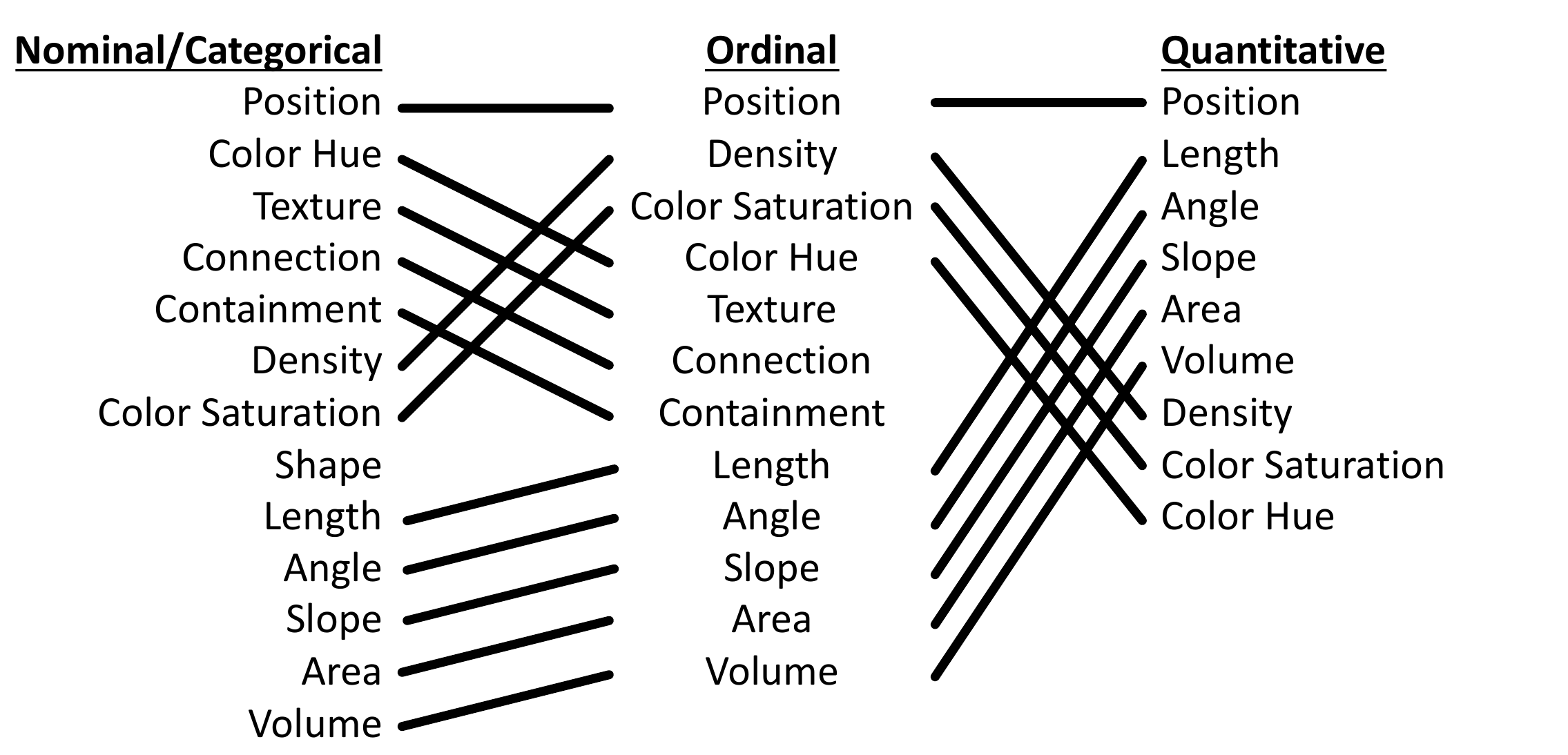}
    \caption{Reproduction of Mackinlay graphical encoding rankings~\cite{mackinlay1986automating}.}
    \label{fig:mackinlay_ranking}
\end{figure}

Understanding the role of perception in the choice of graphical encodings is critical to visualization designers. Based on 10 common graph types---distribution function plots, bar charts, pie charts, divided bar charts, statistical map, curve-difference charts, Cartesian graphs, triple scatterplots, volume charts, and juxtaposed Cartesian graphs---Cleveland and McGill ranked these {perceptual tasks} by the accuracy of quantitative information extraction.  Mackinlay produced one of the earliest comprehensive rankings of graphical encodings by data type, as shown in \autoref{fig:mackinlay_ranking}~\cite{mackinlay1986automating}. The ranking has been further validated and elucidated through followup (replication) studies~\cite{demiralp2014learning, heer2010crowdsourcing, gramazio2014relation, sedlmair2012taxonomy, heer2009sizing, szafir2018modeling, reda2018graphical, neumann1998perception, saket2018evaluating}.

\subsection{Example Replication Studies}

We surveyed major visualization publication venues (IEEE TVCG, ACM CHI, EG EuroVis, and IEEE PacificVis) for papers that cited and then replicated aspects of Cleveland and McGill's work. Our non-comprehensive analysis is primarily limited to the eight papers listed in \autoref{tbl:List_papers}. For each, we provide the objective of the replication study, followed by a high-level description of the novelty added to the replication, which contributed to the visualization research.

\subsubsection{Re-evaluate an Experiment's Objective}

Heer and Bostock's graphical perception study replicated ranking on effectiveness of visual encodings, such as length, position, and angle, using an alternative crowdsourced subject pool on Amazon Mechanical Turk (AMT)~\cite{heer2010crowdsourcing}. Further, they extended the study on additional encodings, such as circular area (e.g., bubble charts) and rectangular area (e.g., treemaps). The ranking order that resulted from their studies were similar to Cleveland and McGill's rankings. The main novelty of this paper was not to validate the findings of Cleveland and McGill, but to test the viability of online user study like crowdsourcing. The results showed that AMT could serve as a viable user study platform for visualization research.

Another replication of Cleveland and McGill was carried out by Saket et al.\ on 12 graphical encodings to study their effectiveness in terms of task completion time and accuracy, when using them for interaction~\cite{saket2018evaluating}. The objective of replication was same as that of original study but re-evaluated on \textit{interactive} graphical encodings. Their ranking followed and confirmed the findings of Cleveland and McGill's, except for a significant difference between length and angle in terms of accuracy.

\subsubsection{Expand an Experiment's Objective}

Hullman et al.~\cite{hullman2011impact} and Harrison et al.~\cite{harrison2013influencing} replicated the study of Cleveland and McGill but ranked their quantitative judgement effectiveness on the basis of social information\footnote{Social information represents the creation and processing of information from multiple features by a group.} and affective priming\footnote{Affective priming is the impact of emotional biases, whose study involves manipulating valence and/or arousal via emotional stimuli.}, respectively. The AMT-based perceptual studies demonstrated {social information} and {affective priming} can significantly influence user's visual judgment~\cite{hullman2011impact, harrison2013influencing}. The findings on social information can be applied to collaborative visualization systems to produce more accurate results on individual interpretations in a social context, and the findings on affective priming showed that it can influence accuracy in common graphical perception tasks.

The concept of graphical perception has extended to various branches of visualization---maps, color, visual properties, etc. Another replication of Cleveland and McGill, towards color mapping on continuous maps, found that spatial frequency significantly impacts the effectiveness of color encodings~\cite{reda2018graphical}. The granular level of novelty to this work is based on conceptual replication of previous work and applying it to a new domain of continuous maps.

A {partial} replication of Cleveland and McGill studied how visual encodings, such as {color, shape and size}, affect a user's way of interpreting data~\cite{demiralp2014learning}. A perceptual kernel\footnote{Perceptual kernel is a distance matrix derived from aggregate perceptual judgments. It contains pairwise perceptual dissimilarity values for a specific set of perceptual stimuli.} is estimated for set of perceptual stimuli, based on size, shape, color, and combinations, to assess the effectiveness of visual representations from reported results of crowdsourced experiments. They compared six stimuli using different set of judgment types---{Likert rating among pairs, ordinal triplet comparisons, and manual spatial arrangements}---to existing perceptual kernels and demonstrated how kernels can be applied to automate visualization design decisions. The novel contribution from this replication work is fixed on similarity judgement using perceptual kernels, extending the concept of the prior work to a particular domain.

\subsubsection{Specialize an Experiment's Objective}

Talbot et al.'s~\cite{talbot2014four} study on bar charts provides insights on comparing adjacent, separated, aligned, and non-aligned bars in types of bar charts. This replication used the foundational work of Cleveland and McGill to focus on perceptual tasks related to bar charts only. {Distractors} (i.e., intermediate bars between bars being compared) affect the comparison between types of bar charts, and inconsistent placement of {marking dots} in the original study affect user accuracy on perceptual tasks.

In similar fashion, the specialized domain replication work of Heer et al.~\cite{heer2009sizing} extended the concept of graphical encoding effect of user judgement to the specialized area of chart height and layering on time and accuracy of a value comparison task. Their findings on estimation accuracy across charts identified transition points in smaller charts, where accuracy and estimation time decreases with size.

For all of the studies mentioned, the additional novelty of the paper helped it stand out beyond just the replication---the studies verified previous results, but they also covered a larger parameter space and/or came to different conclusions than the original study.

\setstretch{1.05}

\section{Discussion}

The overwhelming value of reproducibility is undeniable---from the ability to have third party verification of claims and conclusions, to the development of new insights expanding upon old conclusions, to understanding changes in user expectations and technologies with respect to prior study findings. While general effort toward improved reproducibility have had some success (slow progress but trending in the right direction), replication studies have remained somewhat in the shadows. Further attention needs to be paid to this particular area of reproducibility. One natural avenue is for vision scientists interested in applying their expertise to the area of visualization to contribute replication studies that either re-evaluate, expand, and/or specialize previous studies.

\subsection{You Can't Publish (Strict) Replication Studies}

As long as novelty remains a necessary contribution that reviewers in the visualization community acknowledge, the simple fact is that it will be incredibly difficult to publish strict replication studies. By their nature, the contribution of  strict replication studies is not novelty, thus the reluctance from the reviewers to acknowledge any value. Though some may exist, we did not find any strict replication studies of Cleveland and McGill during our literature survey. \textit{Our opinion is that it remains in the best interests of researchers, in vision science or otherwise, to avoid trying to publish any strict replication study.}

In many ways, the struggles of replication parallel those of application papers in visualization. When application papers are discussed in the halls of the IEEE VIS conference, everyone agrees on their value and wishes there was more acceptance for them. As soon as those individuals review an application paper, a stamp of `limited novelty' (the review equivalent of the `kiss of death') is applied and the paper is promptly rejected. It is only by wrapping the application in novelty that these papers are ever accepted in the main conference tracks. A variety of attempts have been made to correct for this issue, most have been failures. The introduction of \textit{Application Spotlights} this year is the latest attempt, whose success has yet to be determined. The point is, \textit{those looking to promote reproducibility and replication studies should look to the history of application papers for some insights as to what approaches are likely (and unlikely) to succeed moving forward.}

\subsection{How to Publish Replication Studies Anyways}

In this paper, we argue that \textit{the best way for individual researchers to publish replication studies is to distinguish the work with the help of added novelty}. Considering {partial} and {conceptual} style replications bridges the gap between original work and the innovations required for publications. For those new innovations, we demonstrated how prior works used \textit{re-evaluation}, \textit{extension}, and \textit{specialization} to help frame their novel contribution around the replication study, enabling the replication to provide value to the overall contribution of the paper.

\textit{For vision researchers new to visualization, the concept of replication can serve as a bridge into the field}. Replication enables the researcher to become deeply familiar with a specific visualization topic, while contributing to the field. Taking the replication as a base, \textit{re-evaluating}, \textit{expanding}, or \textit{specializing} enables them to contribute added novelty to the field. In this way, the familiarization process (i.e., the replication study) has both personal and community value. For example, re-evaluation work can be used as a foundation for a researcher introducing themselves to visualization community, or an expansion can be used to evaluate conclusions under different conditions, not previously considered. However, we believe specializing represents the best opportunity. Vision scientist can leverage their in deep knowledge of human perception in efforts to replicate and specialize prior visualization studies, thereby bringing innovative ideas to the visualization community. In similar fashion, findings from the vision science community can be replicated and specialized into the context of visualization. In the end, visualization replication studies by vision scientists represent a win-win. First, it engages both communities in a dialog that advances knowledge in both communities. Second, we all benefit when the experimental conclusions we rely on are re-validated and further elaborated upon.

\textit{Replication does not necessarily need new experimentation}. Already completed studies can be utilized to generate new state-of-the-art work. Researchers can make use of available experimental data, study designs, and comparable results from original studies in order to formulate new high-level research questions. A \textit{re-evaluation} of an earlier study can reduce the possibility that conclusions are the result of a statistical fluke, flawed analysis, or a flaw in the study design~\cite{kosara2018skipping}. New perspective on the analysis of that data can be used to \textit{expand} or \textit{specialize} the domain of the work.

Finally, it is important to recognize that many non-replication studies over the years could have corroborated the conclusions of earlier studies by performing small replication studies of key findings. We are not necessarily advocating that all or even the majority of prior studies should be replicated. However, \textit{reviewers could encourage more replication by allocating ``bonus points'' in reviews containing some form of replication.} We are already seeing this with reproducibility in general. In a personal communication with one InfoVis paper chair, they noted a measurably higher average score for papers that included their data in the submission.

\acknowledgments{We would like to thank our reviewers for their valuable feedback on our work. This project is supported in part by National Science Foundation (IIS-1513616 and IIS-1845204).}

\setstretch{0.999}
\bibliographystyle{abbrv-doi}

\bibliography{main}
\end{document}